\documentclass[conference]{IEEEtran}
\IEEEoverridecommandlockouts
\usepackage{cite}
\usepackage{amsmath,amssymb,amsfonts}
\usepackage{algorithmic}
\usepackage{graphicx}
\usepackage{textcomp}
\usepackage{xcolor}
\def\BibTeX{{\rm B\kern-.05em{\sc i\kern-.025em b}\kern-.08em
    T\kern-.1667em\lower.7ex\hbox{E}\kern-.125emX}}
\begin{document}

\title{\title{\huge BlockCampus: A Blockchain-Based DApp for enhancing Student Engagement and Reward Mechanisms in an Academic Community for E-JUST University}
{\footnotesize \textsuperscript{*}Note: Sub-titles are not captured in Xplore and
should not be used}
\thanks{Identify applicable funding agency here. If none, delete this.}
}

\author{\IEEEauthorblockN{Mariam Ayman$^*$, Youssef El-harty$^*$, Ahmed Rashed$^*$, Ahmed Fathy$^*$, Ahmed Abdullah$^*$, Omar Wassim$^*$, Walid Gomaa$^{*a}$}
\IEEEauthorblockA{$^*$Department of Computer Engineering, Egypt-Japan University of Science and Technology, Alexandria, Egypt.}
\IEEEauthorblockA{$^a$Faculty of Engineering, Alexandria University, Alexandria, Egypt}
}
\maketitle

\begin{abstract}
In today's digital age, online communities have become an integral part of our lives, fostering collaboration, knowledge sharing, and community engagement. Higher education institutions, in particular, can greatly benefit from dedicated platforms that facilitate academic discussions and provide incentives for active participation. 
This research paper presents a comprehensive study and implementation of a decentralized application (DApp) leveraging the blockchain technology to address these needs specifically for E-JUST (Egypt-Japan University of Science and Technology) students and academic staff.
\end{abstract}


\section{Introduction}
\label{introduction}

\par 
The Internet is a gateway for students to learn basically anything anytime. The reliance of students on the Internet for information can be quite challenging, not everything is true and/or reliable
which can cause confusion and mistrust as in \cite{9} and \cite{8}. As a result, many platforms were created to provide an online community where people can ask questions and receive answers from a diverse range of users as: Quora \cite{q}, Reddit \cite{r}, and Stack Exchange \cite{s}.
These serve as knowledge-sharing platforms, enabling individuals to seek assistance, share insights, and learn from others' experiences in various fields. They aim to foster collaboration, problem-solving, and knowledge exchange among their users.

\par
However, the problem with these platforms is that they are not foolproof \cite{11}. Maintaining the quality and accuracy of information can be a challenge on these platforms since anyone can contribute answers, there is a risk of incorrect or unreliable information and/or misinformation being shared \cite{12}. Like any online community, these platforms can be susceptible to trolling, spamming, or inappropriate behavior \cite{13}.

\par 
Moreover, over-reliance on individual opinions as answers provided on these platforms are typically based on personal opinions and experiences which is anther critical problem. 
While they can be valuable, they may not always align with expert advice or best practices as shown from \cite{over}. 
Our motivation for creating a dedicated Q$\&$A platform for our university E-JUST poses several challenges. First, the decentralized nature of discussions across different channels creates a fragmented 
knowledge base, making it difficult for users to access relevant information efficiently \cite{21}. Second, the lack of proper incentives and rewards discourages active participation and fails to acknowledge the valuable contributions of the community members \cite{22}. Lastly, the absence of a transparent and immutable system for tracking user reputations limits the ability to establish trust and credibility within the 
community.
The primary objective of this research is to design and implement a blockchain-based DApp, named \textbf{BlockCampus}, that serves as a centralized hub for academic discussions, knowledge sharing, and community engagement within the 
E-JUST ecosystem \cite{10}. 
By leveraging the Ethereum blockchain \cite{00}, we aim to address the aforementioned challenges by introducing a reward mechanism based on user participation and reputation. The DApp will provide a seamless and user-friendly interface, empowering students and academic staff to ask questions, provide answers, vote on content, and earn rewards based on their contributions. The scope of this research paper encompasses the conceptualization, implementation, and evaluation of the DApp, along with the analysis of user engagement and the effectiveness of the reward mechanism \cite{000} \cite{0000}.

\par 
BlockCampus DApp provides a dedicated platform tailored to E-JUST's unique requirements, BlockCampus aims to foster a vibrant and collaborative community, encourage knowledge sharing, and recognize the contributions of individuals within the ecosystem \cite{000}. The blockchain technology ensures transparency, immutability, and decentralization, enhancing trust and enabling a seamless experience for all participants \cite{bitcoin} \cite{000}.
Through this research, we envision a future where students and academic staff at E-JUST (and consequently any other academic institution) 
can harness the power of blockchain technology to fuel knowledge sharing, collaboration, and innovation. By building a vibrant and inclusive community, we strive to create a more enriching and rewarding 
academic experience for all participants. Designed to serve this purpose, BlockCampus is built on top of the Ethereum blockchain and adopts a consortium blockchain model.

\par 
This paper is organized in some sections. Section~\ref{introduction} is an introduction.
Section~\ref{sec: Background} is an overview of the underlying concepts and protocols that were used to make this project possible.
Section~\ref{sec: System Architecture} is an abstract view of the structure of the individual nodes, the blockchain network, and the decentralized app.
Section~\ref{sec: Implementation} is a deep dive into how the technologies the DApp utilizes interact with each other and the blockchain network.
Section~\ref{sec: Testing} is a discussion of the importance of testing.
Section~\ref{sec: Discussion and Conclusion} is a final discussion, some observations, and conclusions. 
Section~\ref{sec: Future Plans} is a discussion of what further improvements can be made and applications that this kind of architecture can be used in.

\section{Background}
\label{sec: Background}

\subsection{Blockchain}
\label{sec: Blockchain}

\par 
The Blockchain technology, initially introduced as the underlying technology for cryptocurrencies like Bitcoin \cite{000}, has gained significant attention in various industries beyond the realm of digital currencies. At its core, \textit{blockchain is a distributed and decentralized ledger} that records and verifies transactions across multiple nodes, providing transparency, immutability, and security. It operates as a chain of blocks, where each block contains a set of transactions, linked together using cryptographic hashes as introduced in \cite{1}.
With the help of these technologies, blockchain provides a secure, reliable, and a decentralized ledger database that can be utilized for many applications and this is greatly shown in \cite{iot} and \cite{3}.

\subsection{Decentralization in Blockchain}
\label{sec: Decentralization in Blockchain}

\par
Centralization is a common form of governance, where we must trust one or few central authorities to maintain order over a structured system, like banks \cite{10}. This creates `Centrality' meaning that if these authorities can't control the structure or entity they operate it will fail, making it vulnerable to attacks and may damage the system or any related entities \cite{bitcoin}. This creates a monopolization of power and a security risk.

\par 
One of the fundamental characteristics of blockchain technology is \textit{decentralization}. Unlike traditional centralized systems that rely on a single authority or intermediary, blockchain distributes data and control among multiple participants (called \textbf{Nodes}), eliminating the need for a central governing entity \cite{b}. This decentralized architecture enhances trust, resilience, and security by removing single points of failure and reducing the risks of failure, manipulation, or censorship \cite{4} \cite{bitcoin}.

\subsection{Ethereum Blockchain}
\label{sec: Ethereum Blockchain}

\par 
Blockchain technology has revolutionized various industries by introducing decentralized and secure systems for data management and transaction processing as indicated by \cite{4} and \cite{2}.
One prominent blockchain platform that has garnered significant attention is \textbf{Ethereum}. It was proposed by Vitalik Buterin in late 2013 and launched in July 2015, with the aim of enabling the development of decentralized applications \textbf{DApps} through the use of smart contracts \cite{10} \cite{e}.

\par 
Ethereum distinguishes itself from traditional blockchain platforms by providing a programmable infrastructure, allowing developers to create and deploy \textbf{smart contracts} on its blockchain \cite{eth}. \textbf{Smart contracts} are self-executing agreements with predefined rules and conditions written into code. They automate the execution and enforcement of contractual obligations without the need for intermediaries, thereby enhancing transparency, efficiency, and trust in various sectors.

\par 
The Ethereum platform operates on a decentralized network of computers, known as nodes, that collectively maintain the integrity and security of the blockchain. Transactions on Ethereum are verified and validated through a consensus mechanism, initially based on proof-of-work (PoW) and transitioning to proof-of-stake (PoS) in Ethereum 2.0 \cite{10}. This consensus mechanism ensures the immutability of the blockchain and prevents malicious activities.
The potential applications of Ethereum extend far beyond simple transactions. Its programmable nature enables the development of complex DApps, security systems, decentralized finance (DeFi) \cite{bitcoin} \cite{0000} protocols, Internet of Things (IOT)\cite{iot} and more. Ethereum has become a breeding ground for innovation and experimentation in the blockchain space, attracting developers, entrepreneurs, and organizations worldwide as it is obvious from.

\begin{figure}
    \centering
    \includegraphics[width=0.5\textwidth]{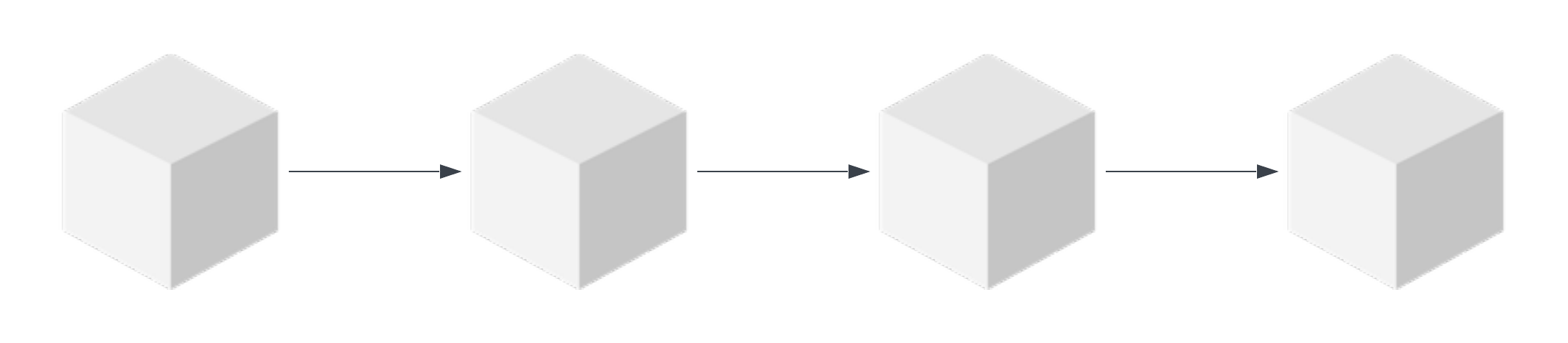}
    \caption{Structure of interconnected blocks}
    \label{fig:my_label}
\end{figure}

\subsection{Smart Contracts}
\label{sec: Smart contracts}

\par 
Smart contracts are \textbf{self-executing} software built on blockchain that automatically facilitate, verify, and enforce the terms of a contract or an agreement without the need for intermediaries. They are programmable contracts that operate on decentralized applications (DApps) and are encoded with specific conditions and actions.
An accurate definition given by Gideon 
Greenspan is “A smart contract is a piece of code which is stored on an Blockchain, triggered by 
Blockchain transactions, and which reads and writes data in that Blockchain’s database”\cite{Green}.
Blockchain transactions are any transfer of data or digital assets on the blockchain network, data could range from reading or writing simple text information to complex smart contract interactions that include asset ownership transfer like tokens and cryptocurrencies.
Once deployed, the smart contract waits for specific triggers or events to occur. These triggers can be predefined conditions, such as a specific date, a certain action by a user, or the fulfillment of certain criteria. When a trigger is met, the smart contract is executed automatically, without the need for any central authority. During execution, the smart contract carries out a series of actions based on its programmed instructions. These actions can include transferring digital assets, updating data records, or triggering other smart contracts. The execution process is entirely deterministic, meaning that the outcome of the contract is predictable and reproducible.

\begin{figure}
    \centering
    \includegraphics[width=0.4\textwidth]{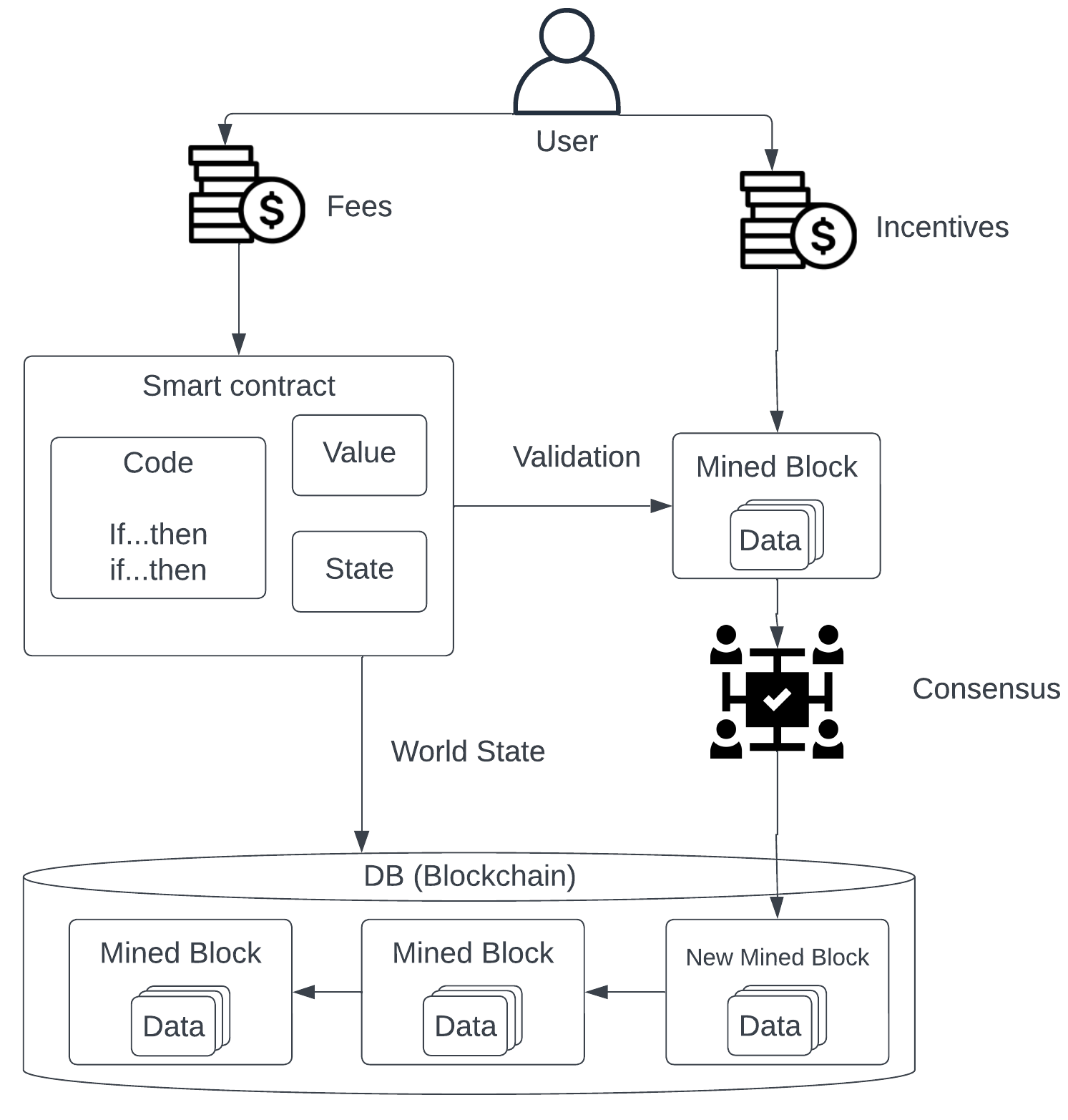}
    \caption{Operational Mechanism of a Smart Contract}
    \label{fig:my_label}
\end{figure}

\par 
Smart contracts offer several notable advantages in DApps. First, they eliminate the need for intermediaries, reducing costs and the potential for human error. By automating the execution and enforcement of agreements, smart contracts enhance the efficiency of transactions and minimize the risk of fraud \cite{eth}.
Moreover, smart contracts provide increased security through their decentralized nature. They are stored on a distributed ledger, making them resistant to tampering and hacking. Once deployed on the blockchain, the terms of the contract are immutable and transparent to all participants, ensuring accountability and reducing disputes.

\par 
Another key benefit of smart contracts is their ability to streamline complex processes. They can handle a wide range of transactions, from simple financial transfers to more intricate multi-step operations. This capability opens up opportunities for various sectors, such as supply chain management, real estate, insurance, and finance, where processes can be automated, and trust can be established through code execution \cite{10}.

\par 
Smart contracts serve as the backbone of DApps, providing a secure, efficient, and transparent mechanism for executing agreements. By removing intermediaries, enhancing security, and streamlining processes, smart contracts have the potential to revolutionize industries and reshape the way transactions are conducted in the digital era\cite{iot}.

\subsection{DApps}
\label{sec: DApps}

\par 
The popularity of blockchain and cryptocurrency has increased in the recent years which gave rise to decentralized Applications (DApps). DApps are an innovative approach to software systems. DApps leverage blockchain technology to provide decentralized, transparent, and secure platforms for various domains. Researchers have explored the potential of DApps in different areas, such as finance, health, and accountability \cite{opp}. These applications offer a range of benefits, including reduced reliance on intermediaries, enhanced data integrity, and increased user control. The surveyed papers shed light on key aspects of DApps, such as their architecture, execution mechanisms, and potential challenges \cite{apps}. They emphasize the importance of accountability and trust in DApp ecosystems and highlight the potential of leveraging blockchain technology to create robust and efficient software systems. By harnessing the power of smart contracts and decentralized networks, DApps offer new avenues for developing innovative solutions, revolutionizing industries, and transforming traditional centralized approaches to software development \cite{opp}.

\section{System Architecture and Logic}
\label{sec: System Architecture}

\subsection{DApp Overview}
\label{sec: DApp overview}

\subsubsection{DApp Goals}
\par
Our DApp, known as BlockCampus, is designed to provide an exceptional online community where users can connect, share knowledge, and interact with peers who share similar interests. Leveraging the power of Ethereum blockchain technology, BlockCampus ensures a reliable and secure platform for students to collaborate and thrive academically. Users of this platform are the stake holders, including students, academic staff (professors and teaching assistants), university administrative staff, developers of the DApp. Each has their own role.

\subsubsection{DApp Users and their Roles}
\label{sec: DApp overview}

\textbf{Students:}
Once they are logged in, students choose their field of interest and join communities with similar interests. Upon completion, they would have the ability to view the community posts, questions and interact with other members of the community. An Example of these interactions is posting questions or inquiring about topics related to the community and sharing useful study materials, such as asking the proof a mathematical problem or asking questions concerning a programming language or advancements in finance and business models. These questions can be commented on, which is the primary way for users to interact with each other. Comments and posts can be voted on by other users and along with other parameters, they are ranked accordingly, and the most rated contributions are shown first.
\textbf{Academic staff:}
In addition to the features mentioned above, the academic staff members on BlockCampus play a pivotal role in fostering an engaging and productive learning environment. Academic staff have access to an expanded set of functionalities that allow them to not only participate in the community but also contribute to its growth and development. They possess additional privileges and responsibilities that enhance their engagement within the community. One of which is the ability to rate students’ contributions with their distinct rating system that provides expert feedback and guidance to students' queries. They can comment on posts, provide clarifications, suggest alternative approaches, and offer constructive criticism when necessary.
\textbf{EJUST administrative staff:}
Since the DApp is built in a decentralized manner, users have complete freedom to express their thoughts, opinions, and ideas openly in the platform which is one of the main features of the DApp. But sometimes these opinions can be controversial and sometimes inappropriate and unrelated to the community, and in some cases lead to harassment and trolling. So, to ensure a safe and conflict free environment, community guidelines that are compatible with the university rules are implemented to mitigate illegal activities, spread of false information, and introduce a culture of respect and inclusivity.
The role of administrative staff is to do exactly that by moderating the overall DApp content and flagging inappropriate content while providing freedom in opinion at the same time.
\textbf{Developers:}
Their role is the most important as they keep the operation, maintenance, and updates of the system throughout the lifecycle of the DApp. Having a well maintained DApp creates a secure and reliable system invulnerable to attacks and safeguards the community. This ensures optimal user experience, development and growth of the platform.

\subsubsection{Incentives: Reputations System and Awards}
\label{Incentives: reputations system and awards}

\par
Rewards can help foster long-term engagement and retention within the community. Users who consistently contribute and earn rewards feel a sense of loyalty and commitment to the community. This sustained engagement promotes a stable and active user base, ensuring the continuity and growth of the community over time.

\textbf{Reputation points (Bateekh):}
As a way to recognize contributions and establish a sense of trust between users, a reputation system was made to measure credibility within the community and it’s a reflection of the user’s engagement, behavior, trustworthiness, and commitment. The simplest way to implement such a system is by using points, these points are affected by the engagement to the contribution through comments, votes, and rewards (more on them later). Each upvote increases the points of the user based on precise calculations that takes into account the rating of the staff, reputation of the user, and the age of the post. Combined, they give an accurate estimate of how mush points the user should earn, the higher these parameters are the greater the points.
Users with enough reputation points would earn the native token of the DApp (Tofu) that can be used on the platform. These tokens can be used in exchange for wavering university tuition fees, bonus grades, or services provided by other users or the university. Examples on the services are free courses, premium features, customizing the profile, and gain exclusive material. These services can vary depending on the organization and the user, that is the academic staff may have services available to them that are different then the students.
The distribution of the tokens and points is managed automatically by the Token smart contract and the distribution model created within the contract by the developers. More details are discussed in Section 3.4 and 3.5. 

\textbf{Awards:} 
Another fun way to show appreciation and gratitude to users is rewards. They make the users feel recognized and valued for their efforts. This recognition acts as a motivator, encouraging them to continue participating actively.
As an incentive, user who receive recognition and Awards for their contributions are more likely to attract attention from other students and academic staff. This can lead to networking opportunities, mentorship possibilities, or even career advancements for the rewarded users.

\subsection{Blockchain Architecture}
\label{sec: Blockchain Architecture}

\subsubsection{Block Anatomy}
\label{sec: Block Anatomy}

\par 
The blockchain in the BlockCampus DApp consists of a series of blocks as seen in Figure [1], and each block contains a set of transactions. These transactions can represent various activities within the DApp, such as posting questions, providing answers, voting on content, or earning rewards. Additionally, each block is linked to the previous block through cryptographic hashes which serve as a unique identifier for each block and contains the information of the previous block's hash. This linking mechanism creates a chain of blocks, ensuring the integrity and immutability of the blockchain Figure [3]. If any modification is attempted on a block, it would result in a change in its hash, which, in turn, would affect the subsequent block's hash. This property makes it extremely difficult for malicious actors to tamper with the blockchain, as any tampering would be immediately detectable \cite{bitcoin}.

\begin{figure}
    \centering
    \includegraphics[width=0.4\textwidth]{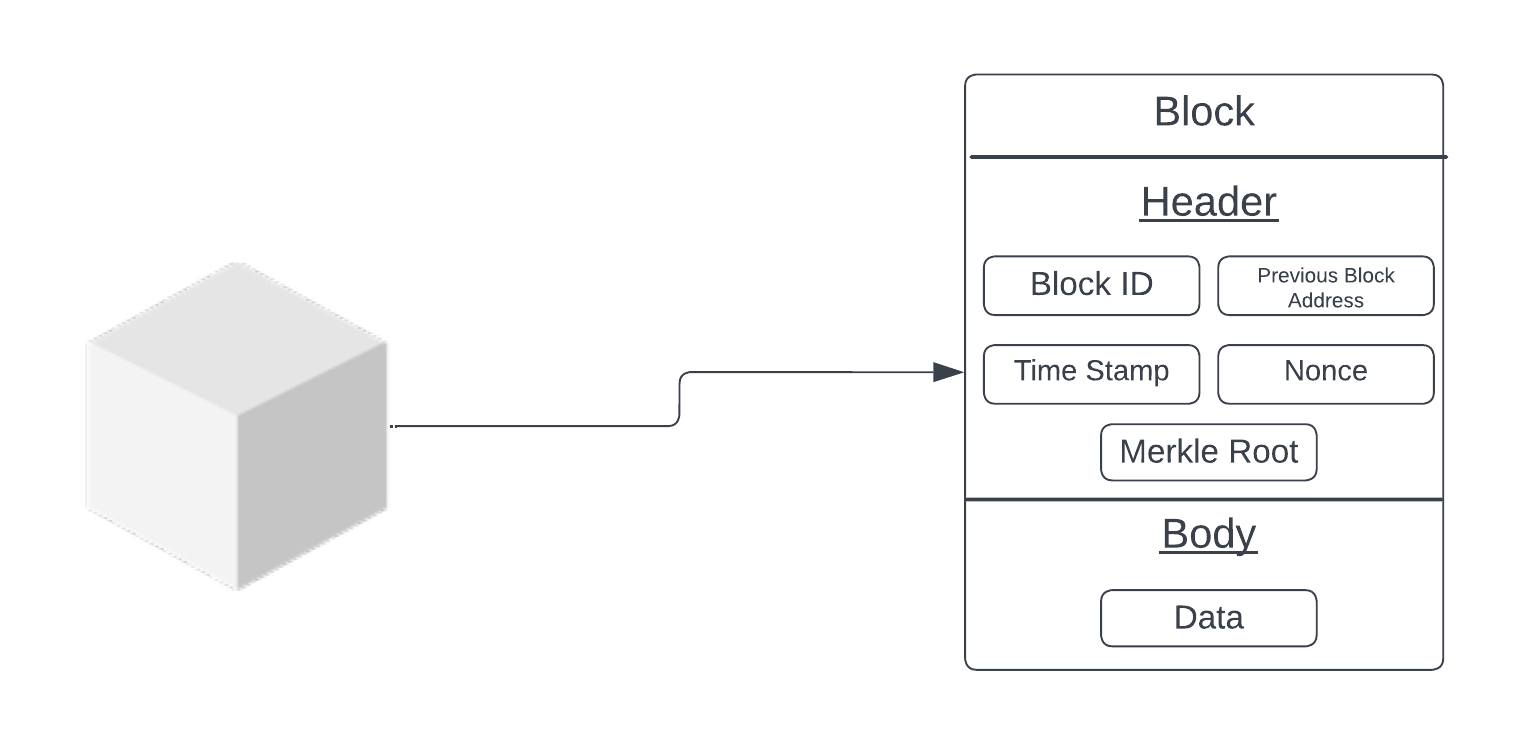}
    \caption{Block Anatomy}
    \label{fig:my_label}
\end{figure}

\subsubsection{Blockchain Type (Consortium)}
\label{sec: Blockchain Type (Consortium)}

\par 
BlockCampus adopts a \textit{consortium blockchain model} to meet the specific requirements of the E-JUST ecosystem. Unlike public blockchains, which are open to anyone for participation and validation, a consortium blockchain restricts access to a predefined set of trusted entities.
In the case of BlockCampus, the consortium members include E-JUST students, academic staff, and other relevant stakeholders within the university ecosystem. By limiting the participation to trusted entities, the consortium blockchain enhances privacy, scalability, and transaction throughput compared to public blockchains. Therefore, the participants have a higher level of control and can make decisions collectively, enabling more efficient consensus mechanisms. It also offers enhanced privacy as the transactions and data shared within the consortium are not visible to the public. This privacy is particularly important for academic discussions and sensitive information sharing within the E-JUST community. In addition, it ensures a higher degree of trust among participants since they are known entities within the ecosystem which facilitates effective collaboration, knowledge sharing, and community engagement.
By utilizing a consortium blockchain, BlockCampus can strike a balance between openness and privacy, allowing the E-JUST community to benefit from the advantages of blockchain technology while maintaining control over the network and data.

\subsubsection{Node Architecture}
\label{sec: Node Architecture}

\par 
The BlockCampus blockchain network consists of multiple nodes, which can be represented by E-JUST, academic staff, students, and other authorized stakeholders. These nodes are distributed across different locations and are connected through the blockchain network. Each node maintains a copy of the blockchain and participates in the consensus process.
The distributed nature of the node architecture ensures that no single entity or centralized authority has complete control over the network. This decentralization enhances the security and resilience of the network, as it becomes more resistant to single points of failure and external attacks. If one node goes offline or becomes compromised, the other nodes in the network continue to validate transactions and maintain the integrity of the blockchain.

\subsubsection{Consensus Mechanism (PoA)}
\label{sec: Consensus Mechanism (PoA)}

\par 
To maintain consensus on the state of the blockchain and validate transactions, BlockCampus employs a proof-of-authority (PoA) consensus mechanism. In PoA, consortium members take turns acting as validators based on their authority within the network.
In the BlockCampus DApp, the consortium members are E-JUST administrative staff (as regulators and compliance authorities).They are given the authority to validate transactions. Each validator takes turns adding blocks to the blockchain and confirming the validity of transactions. This consensus mechanism ensures faster transaction confirmation times and higher throughput compared to more computationally intensive consensus mechanisms like proof-of-work (PoW).
The applied PoA consensus mechanism is suitable for our consortium blockchains as it strikes a balance between decentralization and scalability. Consortium members are known and trusted entities within the network and their authority allow for efficient transaction validation and consensus.

\subsection{DApp Architecture}
\label{sec: DApp Architecture}

\begin{figure}
    \centering
    \includegraphics[width=0.4\textwidth]{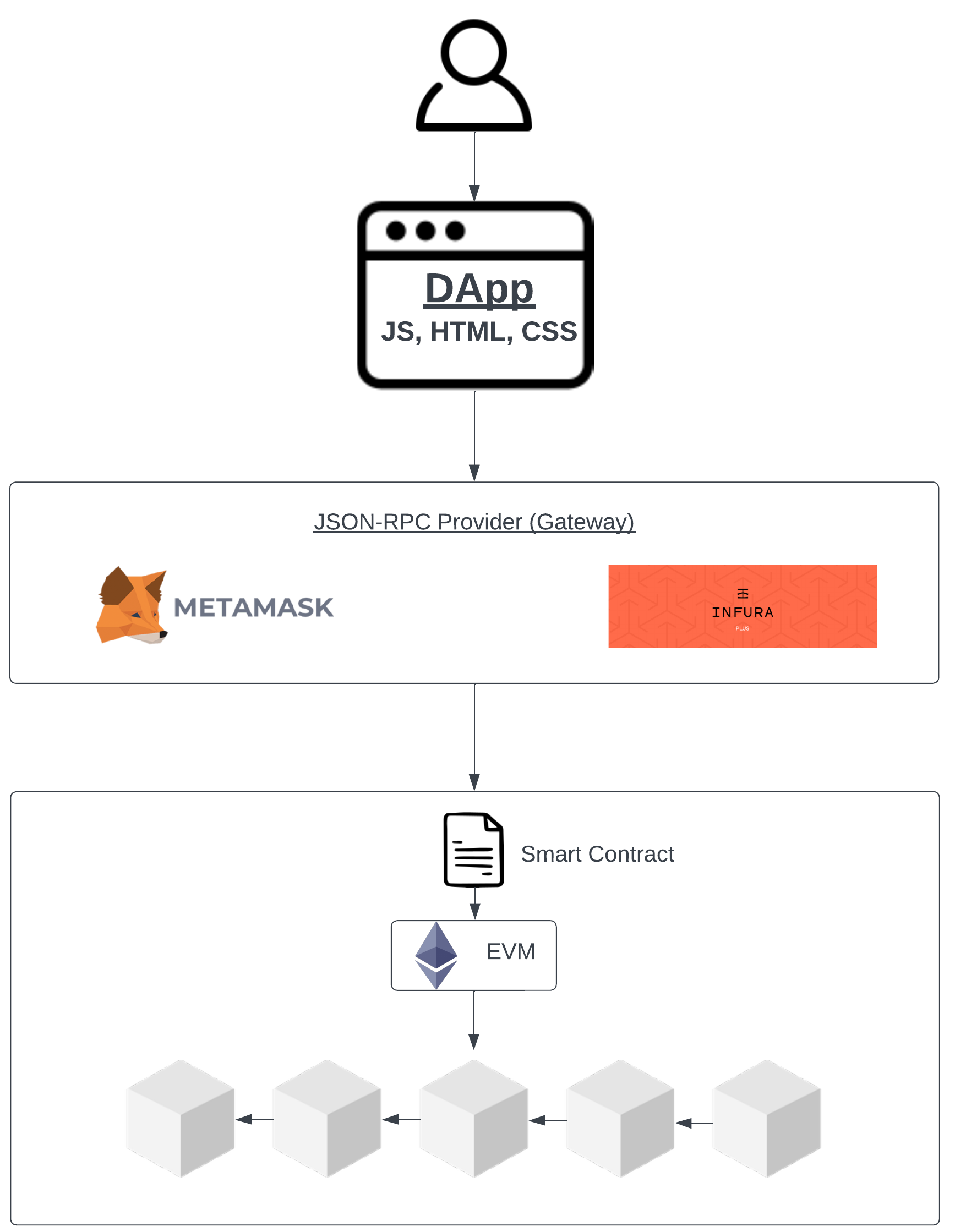}
    \caption{DApp Architecture}
    \label{fig:my_label}
\end{figure}

\par 
As shown in Figure [4], the BlockCampus DApp architecture consists of the following key components:

\par
\textbf{User Interface (UI) Layer:} The UI layer is responsible for providing an intuitive and user-friendly interface for users to interact with the BlockCampus DApp. It includes the graphical user interface (GUI) elements such as web pages, forms, buttons, and menus that enable users to access different functionalities of the DApp. The UI layer is designed to be responsive, ensuring a seamless experience across different devices and screen sizes.

\par
\textbf{Backend Layer:} The backend layer in the BlockCampus DApp architecture plays a pivotal role in facilitating seamless interactions between the user interface layer, smart contracts, and IPFS (InterPlanetary File System) integration. Web3 serves are used as a bridge between these components, handling communication between the DApp's UI and the blockchain network.
One of the key responsibilities of the backend layer is to interact with the smart contracts deployed on the Ethereum blockchain. Through the API calls of Web3 and the integration with the smart contracts written in Solidity, the backend layer enables the execution of transactions and the retrieval and updating of data stored on the blockchain. This ensures transparency, security, and immutability of the information exchanged within the BlockCampus ecosystem.

\begin{figure}
    \centering
    \includegraphics[width=0.5\textwidth]{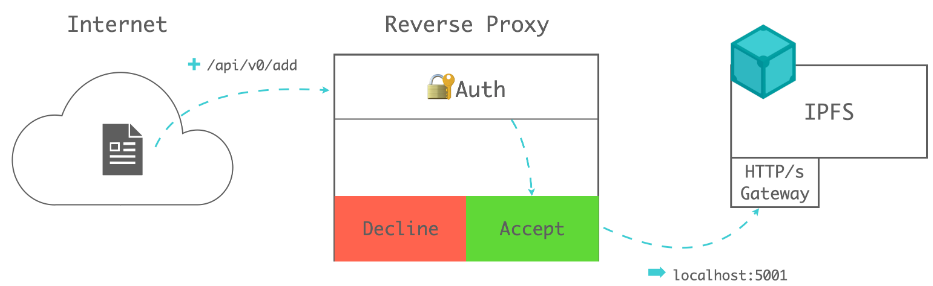}
    \caption{IPFS Gateway (Adapted from \cite{ipfs-g})}
    \label{fig:my_label}
\end{figure}

\par 
As shown from figure[5], the layer handles all the critical functionalities of the logic of the DApp as: safe data storage and communication with the blockchain network, including interactions with smart contracts and leveraging IPFS functionalities for decentralized file storage and retrieval. Additionally, the backend layer enforces access controls and ensures user authentication and authorization, safeguarding sensitive information and maintaining user privacy within the BlockCampus ecosystem.

\par 
\textbf{Blockchain Network:} The blockchain network forms the foundation of the BlockCampus DApp architecture. In our case, the DApp is built on top of the Ethereum blockchain which comprises multiple nodes that collaborate to maintain the integrity and security of our blockchain.

\par 
The interactions within the BlockCampus DApp architecture follow a coordinated flow as shown from figure [6]. Users interact with the user interface layer, accessing various functionalities and submitting requests. 
These requests are then processed by the backend layer, which communicates with the Ethereum blockchain network, including IPFS, to execute transactions, store files, and retrieve data from smart 
contracts and IPFS storage. The backend layer handles the necessary computations, validates inputs, and ensures data integrity and security. Finally, the results are processed by the backend layer and presented to the user through the user interface layer, enabling a seamless and secure user experience within the BlockCampus DApp.
\par 
Overall, the DApp architecture combines the user-friendly interface, the logic and rules defined in smart contracts, the backend layer for processing and communication, and the underlying blockchain network to provide a decentralized, transparent, and secure platform for academic discussions, knowledge sharing, and community engagement within the E-JUST ecosystem. 

\begin{figure}
    \centering
    \includegraphics[width=0.5\textwidth]{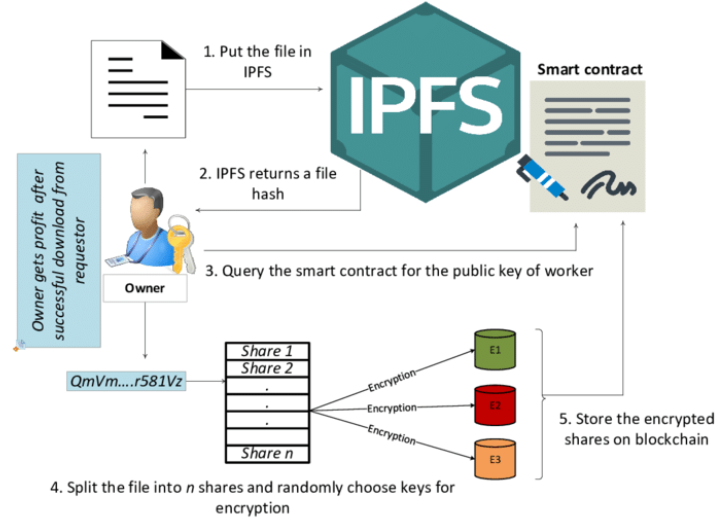}
    \caption{Data Sharing on IPFS (Adapted from \cite{ipfs})}
    \label{fig:my_label}
\end{figure}

\bigskip

\subsection{Tokens Creation}
\label{sec: Tokens Creation}

\par 
In order to establish a vibrant ecosystem within our blockchain platform, the creation of tokens plays a crucial role. We have implemented a robust token creation process that adheres to industry standards and best practices. Our platform supports token standards  as ERC-20, which enable seamless integration with existing decentralized applications (DApps) and ensure compatibility with various wallets and exchanges.
To create our tokens, we utilize smart contracts that facilitate the automatic issuance and management of tokens, ensuring transparency and security. Through the use of smart contracts, we eliminate the need for intermediaries and enable peer-to-peer token transfers.

\subsection{Tokenomics}
Tokenomics forms the foundation of our token's value proposition and economic model. By carefully designing the tokenomics, we aim to incentivize adoption, promote network growth, and reward DApp users.
The token's value is influenced by factors such as supply and demand dynamics, utility within the ecosystem, and the overall economic model. Additionally, we have incorporated mechanisms that recognize and reward user contributions to the community:

\textbf{Bateekh:} Bateekh is the reputation point of our system, indicating a user's contribution to the community through questions and answers. Users will gain Bateekh when others upvote their questions or answers. The rating of the academic staff will also grant Bateekh points to users. The number of points gained depends on various parameters, including the age of the questions or answers and the rating of the academic staff.

\textbf{Tofu:} Tofu serves as the main currency (Token) of our website and is utilized in all transactions. It holds tangible value as users can exchange Tofu for services on the website or receive acknowledgment from the university. The Tofu token serves as a reward mechanism within our ecosystem, further motivating community engagement and participation.

\subsubsection{Distribution Model}
\label{sec: Distribution Model}

\par 
The distribution model refers to the way in which tokens are initially allocated and distributed among users or participants. It determines how the token supply is divided and distributed among various stakeholders.
In our application we used the following distribution model:

\par 
\textbf{- Token supply:} With accordance to the ERC-20 standards for token creation and the project's goals, we decided to launch the token with an initial supply that would be enough to reward users for their contribution without overflowing the DApp with too much tokens. And the max token supply will be fixed and determined to ensure a predictable and finite token availability.

\textbf{- Token utility:}
Tofu token will exclusively serve as a reward mechanism within the DApp ecosystem. Users can earn Tofu tokens by actively participating in the community, contributing valuable content, providing insightful comments, or receiving positive feedback from other users.

\textbf{- Token Rewards:} Users who accumulate Tofu tokens can redeem them for various services offered within the DApp. These services can include accessing premium content, receiving personalized assistance or support, obtaining exclusive features, or participating in specialized activities or events.

\textbf{- Token Exchange:} Tofu tokens will not be tradable on external cryptocurrency exchanges. Instead, users can only use the tokens within the DApp to access the designated services. This ensures that the tokens retain their value within the ecosystem and encourages users to actively participate to earn more tokens.

\textbf{- Token Burns}: To maintain the value of Tofu tokens, a portion of the tokens may be periodically burned or permanently removed from circulation. This can be done by the DApp's smart contract, ensuring that the token supply is stable over time, potentially keeping the value of the remaining tokens constant. In some cases where a user stops using the DApp and deactivates his account, or the user is no longer affiliated with the university, his/her tokens will be burned as to not keep unused tokens in the network.

\textbf{- Token Grants:} 
Tokens are allocated to specific individuals, organizations, or community members as a reward, incentive, or support for their contributions to the project. The distribution model chosen for a token project depends on factors such as the project's goals, regulatory considerations, funding needs, community engagement strategy, and desired token economics.
Therefore, to ensure a fair and transparent token distribution, we have devised a comprehensive distribution model based on the proposed model. The allocation of tokens is designed to benefit different stakeholders involved in the project, including investors, team members, advisors, and community members. We have allocated tokens based on predetermined percentages to ensure equity and align incentives.
Furthermore, the distribution model takes into account the contributions of community members through the accumulation of Tofu tokens. Users who actively contribute and accrue Tofu tokens may be eligible for additional token allocations or rewards, fostering a sense of fairness and recognition within the community.

\subsubsection{Access Control}

\par 
Access controls refer to the mechanisms and rules in a smart contract that govern who can perform certain actions or access specific functions or data within the contract. Access controls are implemented through a set of defined functions and rules written in the smart contracts to ensure that only authorized individuals or entities can execute specific operations, modify contract state, or access sensitive information. Access controls won’t affect the decentrality of the network as they only affect sensitive information that only developers and owners of the blockchain are interested in.
They are crucial for maintaining the security, integrity, and proper functioning of a smart contract. They help prevent unauthorized actions or malicious behavior that could compromise the contract or its associated assets. By enforcing access controls, you can mitigate risks and ensure that only trusted parties can perform critical operations or access sensitive data.

\par 
In addition to the ownership role, We implement the Role-Based Access Control (RBAC) which assigns specific roles and privileges to certain user accounts, such as Teaching Assistants (TAs) and Professors as they have the authority to rate and validate answers provided by all users. This ensures that contributions from these designated individuals hold significance and are recognized within the BlockCampus community.
Through implementing access control mechanisms using the smart contracts, we provide a secure and controlled environment for users, ensuring that only authorized individuals can perform specific actions or access certain features of the DApp which helps maintain the integrity of the system and prevents unauthorized use or malicious activities.

\subsection{User Management and Authentication}
\label{sec: User Management and Authentication}

\par 
User management and authentication are integral parts of the BlockCampus DApp, ensuring secure access and personalized experiences for users. The system supports two main user categories: students and academic staff. \textbf{For students,} functions such as registration, login, asking questions, answering questions, voting, giving awards, and posting are available. Students' reputation, known as Bateekh, is influenced by their question and answer contributions. \textbf{For academic staff members,} including professors and teaching assistants, they have similar functionalities to students, but with additional information during registration, such as their academic staff ID and position. User management and authentication are facilitated through the use of usernames and passwords, allowing users to securely access their accounts.
On top of that, the blockchain provides a unique address to each user, and this address is included in all transactions within the network. This creates an extra layer of security as any fraudulent activity such as tampering with transactions or data in the blockchain can be detected and traced back to the attacker who will be penalized accordingly. 
By implementing robust user management and authentication mechanisms, BlockCampus ensures the privacy and security of user information, while providing a seamless and tailored experience for both students and academic staff.

\subsection{Rewarding Mechanism}
\label{sec: Note}

\par
As the project get into the production and expands, the rewarding mechanism or token distribution can be adjusted based on the needs and feedback received from the platform's users and stakeholders. Launching provides an opportunity to gather real-world data and insights on user engagement, behavior, and preferences. This information can be used to refine and optimize the rewarding mechanism and token distribution to ensure its effectiveness and alignment with the platform's goals.
Users may provide feedback on the current rewarding mechanism, suggesting improvements or highlighting areas where adjustments could enhance their experience. Analyzing user behavior and engagement patterns can also benefit the developers by providing information about the effectiveness of the current rewarding mechanism. It may reveal areas where adjustments can be made to better incentivize desired user actions or promote specific behaviors.
As the platform grows and attracts a larger user base (if, for example, other universities joins the platform), the initial token distribution model may need to be adjusted to accommodate the increased demand for rewards. Scaling the rewarding mechanism ensures that the platform can sustainably support a growing user community.

\section{Implementation}
\label{sec: Implementation}

\subsection{Blockchain Network}
\label{sec: Blockchain Network}

\par 
One of our goals was to create a private blockchain network to deploy the smart contracts on. Creating a private Ethereum network built on Besu Hyperledger involves setting up a local blockchain environment that allows for secure and private transactions within a specific network of participants. This private network can be utilized for various purposes such as testing and development of our DApp.

\par 
To create a private Ethereum network with Besu Hyperledger, several steps need to be followed. Firstly, the network participants must set up the required infrastructure, including installing the Besu client and configuring the network parameters. Besu is an open-source Ethereum client developed by Hyperledger that offers enhanced privacy features and enterprise-grade capabilities.

Once the infrastructure is in place, the network's genesis block needs to be defined. The genesis block serves as the initial state of the blockchain and includes information such as the network ID, chain ID, account balances, and other network-specific configurations. It can be customized to meet the specific requirements of the private network.

Next, the network participants need to configure the nodes to join the private network. Each participant should have a node running the Besu client and connect to the network using the appropriate network ID and bootnodes. Bootnodes act as the initial network entry points and help nodes discover and connect with each other.

Once the nodes are connected, the private network is ready for use. Participants can interact with the network by deploying and executing smart contracts, sending transactions, and accessing the blockchain's data. The private nature of the network ensures that only authorized participants can access and participate in the blockchain's activities.

By creating a private Ethereum network with Besu Hyperledger, we can leverage the benefits of blockchain technology while maintaining control over the network's privacy and configuration. This allows for experimentation, development, and deployment of decentralized applications in a secure and controlled environment, suited for the specific needs of the participants.

\subsection{TechTools}
\label{sec: TechTools}

\par 
The development of the DApp involved the utilization of various tools and technologies to ensure efficient and secure implementation. This section provides an overview of the key tools and technologies employed throughout the development process.

\textbf{Hardhat: }A popular development framework for Ethereum, was chosen as the primary tool for compiling, testing, and deploying smart contracts. Hardhat offers a comprehensive suite of developer-friendly features, including built-in testing capabilities, a robust plugin system, and integration with popular Ethereum networks. Its extensive functionality and extensive community support made it an ideal choice for the development workflow.

\textbf{React: }For the front-end application, the React framework was employed due to its flexibility, modularity, and the extensive ecosystem of libraries and tools. React enables the creation of interactive user interfaces, providing a smooth and responsive user experience. Additionally, its component-based architecture facilitates code reusability and maintainability, contributing to efficient development practices.

\textbf{Remix IDE: }Remix IDE, a web-based integrated development environment (IDE), played a vital role in testing the smart contracts. With its intuitive interface and powerful debugging capabilities, Remix IDE enabled developers to deploy and interact with contracts in a simulated environment. Its comprehensive testing framework facilitated thorough unit testing and ensured the reliability and correctness of the smart contract code.

\textbf{Besu Hyperledger: }To establish a private blockchain network, Besu Hyperledger was utilized. Besu, an Ethereum client developed by Hyperledger, offers enterprise-grade features and enhanced privacy mechanisms. It 
enables the creation of permissioned networks with configurable consensus algorithms and network parameters. By leveraging Besu Hyperledger, the DApp ensured a secure and controlled environment for transaction processing 
and data storage.

\textbf{MetaMask: }MetaMask, a widely adopted wallet and browser extension, served as the primary tool for user wallet management and interaction with the DApp. MetaMask provided a convenient and secure way for users to access and interact with the Ethereum network, sign transactions, and manage their digital assets. Its compatibility with multiple browsers and user-friendly interface made it a popular choice among Ethereum users.

\textbf{Infura: }Infura played a crucial role in providing reliable and scalable access to the Ethereum Virtual Machine (EVM). By leveraging Infura's RPC endpoint, the DApp was able to connect to the Ethereum network without the need to run a local node. Infura's infrastructure allowed for seamless deployment and interaction with smart contracts, ensuring high availability and reliability.

The development of the DApp involved a comprehensive stack of tools and technologies. The combination of these tools and technologies along with the features and services of the blockchain technology formed a robust foundation for the development and deployment of the DApp.

\subsection{Challenges and Optimizations}
\label{sec: Challenges and Optimizations}

\par 
Throughout the development process, several challenges were encountered, primarily stemming from the deprecation of certain frameworks and tools. These deprecations posed significant hurdles that required careful mitigation strategies to overcome. One notable challenge encountered was the deprecation of various frameworks and tools that were initially relied upon for certain development tasks. For instance, some testnets were discontinued which made the conversion to other testing environments a necessity. Another example for testnets, Georli testnet (which is the most common testnet) only provided test ether for users who have real Ether in the Ethereum Mainnet, which was a recent decision due to the abuse of users to the Georli faucets. This migration resulted in compatibility issues and error-prone deployments. This required the exploration and adoption of alternative tools and approaches to accomplish these tasks effectively.

\par The deprecation of Microsoft Azure Blockchain-as-a-Service (BaaS) in 2021 also posed a significant challenge in creating a private blockchain network and hosting it on the cloud. Azure BaaS provided a convenient and streamlined platform for deploying private blockchains, but its deprecation necessitated the exploration of alternative cloud providers and infrastructure solutions. We explored ConsenSys Quorum blockchain service which was very similar to Azure's BaaS, but like its counterpart, it was deprecated on march 2023. So a decision to use Sepolia testnet was made.

\par 
As with any complex development project, the process was not without its fair share of errors and troubleshooting. The integration of multiple tools and frameworks introduced potential points of failure and compatibility issues. Addressing these errors required meticulous debugging, error handling, and collaboration among the development team. Thorough testing and meticulous code review were essential in identifying and resolving these issues promptly.

The development of the DApp involved the utilization of various tools and technologies that were relatively new to the development team. This introduced a steep learning curve and necessitated acquiring new knowledge and skills to effectively leverage these tools. To comprehend the underlying mechanisms of blockchain technology and the secure nature of Ethereum, the development team embarked on a journey of studying cryptography. Understanding concepts such as encryption, digital signatures, and cryptographic hashing algorithms was crucial to grasp the security principles of blockchain and effectively implement them within the DApp. The learning process involved rigorous self-study, consultation of academic resources, and engaging in online courses to gain a solid foundation in cryptography and blockchain fundamentals.

\section{Testing}
\label{sec: Testing}

\par 
Blockchains are immutable, no data within the network can be changed or deleted, that’s one of the key features of the blockchain technology. So, modifying any data would be very difficult and 
require the consensus of the whole network. For these reasons, testing a smart contract before deploying is a requirement to ensure that the contract satisfies requirements for reliability, 
usability, and security. An ideal approach of testing was using a local Ethereum environment that would ease the process of testing and catching any minor or major flaws in the smart contract. 
Remix IDE, a tool that is used for testing the contracts locally without affecting the mainnet, is used in our case, it provides all the tools needed for a comprehensive testing, and a 
compatible environment focused on handling and discovering bugs within the system. Testing is a crucial step in development, upgrade or editing in the blockchain, if it’s withing the scope of 
possibility, would result in major errors and can cause massive losses for users.
Before writing or developing the smart contract, in-depth research into the working mechanism of the smart contracts and the underlying blockchain was done. As well as understanding the 
requirements of the DApp, how users will access and use those functionalities, assumptions regarding contract execution and conducting unit tests to validate those assumptions and implementing 
any security or reliability measures necessary for a reliable service. This is particularly useful for running happy path tests that determine if functions in a contract return the correct 
output for valid user inputs. Another effective approach is to go beyond conducting tests for expected user behavior and include negative tests that assess the failure of functions and how users 
can manipulate or abuse the system for their favor (Cases such as cheating or creating fake accounts to upvote their own posts and comments).

\section{Discussion and Conclusions}
\label{sec: Discussion and Conclusion}

\par 
We presented the design, implementation, and evaluation of the BlockCampus DApp, a blockchain-based decentralized application tailored for E-JUST university students and academic staff. 
The DApp serves as a centralized hub for academic discussions, knowledge sharing, and community engagement within the E-JUST ecosystem.

\par 
By leveraging the Ethereum blockchain, the BlockCampus DApp addresses the challenges of fragmented knowledge base, lack of incentives, and transparency in tracking user reputation. The blockchain architecture ensures transparency, immutability, and decentralization, enhancing trust and security. The consortium blockchain model, with trusted entities as participants, provides privacy, scalability, and transaction throughput while maintaining control over the network and data. The PoA consensus mechanism ensures fast transaction confirmation times and higher throughput.

\par 
The token creation and tokenomics model incentivize adoption, promote network growth, and reward user's contributions. Bateekh and Tofu tokens provide tangible value within the ecosystem, recognizing and rewarding user participation and fostering community engagement.

\par 
The implementation of the BlockCampus DApp has the potential to foster a vibrant and collaborative community within the academic environment. By harnessing the power of blockchain technology, E-JUST students and academic staff can benefit from enhanced knowledge sharing, collaboration, and innovation. The transparent and decentralized nature of the DApp promotes trust and facilitates a rewarding academic experience for all participants. Future enhancements to the BlockCampus DApp could include further refining the tokenomics model, expanding the functionality and user base, and integrating additional features to enhance user experience and engagement.

\par 
In conclusion, the BlockCampus DApp presents a promising solution to enhance community engagement, knowledge sharing, and collaboration within the academic settings of E-JUST. By leveraging the 
blockchain technology and implementing a robust tokenomics model, the DApp offers transparency, incentivization, and trust among participants. However, further research, evaluation, and refinement are needed to fully realize the potential of the BlockCampus DApp and its impact on academic communities. 

\section{Future Plans}
\label{sec: Future Plans}

\par 
Our project opens up avenues for future research and exploration. Additional studies could focus on the scalability of the DApp architecture, exploring solutions such as sharding or layer-two protocols to accommodate a growing user base and increasing transaction volume \cite{222}. Security audits and vulnerability assessments can be conducted to ensure the robustness and resilience of the DApp against potential attacks.

\par 
Moreover, extending the functionality of the BlockCampus DApp to include features such as collaborative document editing, peer-to-peer mentoring, or research collaboration tools could enhance its value proposition and attract a wider range of users. Integration with other academic systems and platforms, such as learning management systems or academic repositories, can further streamline the academic experience and facilitate seamless knowledge sharing.

\par 
The utilization of blockchain-based decentralized applications (DApps) also has the potential to revolutionize the way students are granted ownership of their credentials, such as certificates and diplomas. By harnessing the power of blockchain technology, it becomes possible to ally this innovative approach with education institutions, making the process of verifying academic qualifications more efficient, streamlined, and transparent.

\par 
Traditionally, verifying academic credentials involves complex and time-consuming procedures, relying on manual checks and third-party intermediaries. However, by leveraging the blockchain technology, a tamper-proof and decentralized ledger, the entire verification process can be facilitated. Each academic achievement can be securely recorded on the blockchain, providing an immutable record of a student's accomplishments. With blockchain-enabled DApps, employers, educational institutions, and other stakeholders can effortlessly authenticate the validity of a student's credentials, eliminating the need for cumbersome paperwork and reducing the chances of fraudulent qualifications. Moreover, this system ensures that students retain ownership and control over their academic records, empowering them to share their achievements securely and seamlessly with potential employers or academic institutions worldwide.

\par 
The integration of blockchain technology into the realm of education not only enhances the trust and reliability of academic qualifications but also fosters a more accessible and inclusive environment for learners.

\section*{Acknowledgements}
\label{sec: Acknowledgements}
We would like to express our sincere gratitude to Dr. Walid Gomaa for his invaluable guidance, expertise, and support throughout the entire research process of this paper. Dr. Gomaa's profound knowledge in the field, as well as his insightful feedback and constructive criticism, have significantly contributed to the quality and depth of this research.

Additionally, we would like to express our gratitude to Eng. Amine Ben Mansour for his valuable contributions and assistance throughout the project. His expertise, collaboration, and dedication played a crucial role in the successful completion of this research. His experience in Blockchain development enabled us to achieve our research goals.

\bibliographystyle{IEEEtran}
\bibliography{ref}

\end{document}